# High-performance Thermal Interface Material Based on Few-layer Graphene Composite

*Wonjun Park[1, 2] *, Yufen Guo[5], Xiangyu Li[3, 2], Jiuning Hu[6], Liwei Liu[5], Xiulin Ruan[3, 2], Yong P. Chen[4, 2, 1] **

[1]School of Electrical and Computer Engineering, Purdue University, West Lafayette, IN 47907, U.S.A.

[2]Birck Nanotechnology Center, Purdue University, West Lafayette, IN 47907, U.S.A.

[3]School of Mechanical Engineering, Purdue University, West Lafayette, IN 47907, U.S.A.

[4]Department of Physics and Astronomy, Purdue University, West Lafayette, IN 47907, U.S.A.

[5]Key Laboratory of Nanodevices and Applications, Suzhou Institute of Nano-Tech and Nano-Bionics, Chinese Academy of Sciences, Suzhou, Jiangsu 215123, People's Republic of China.

[6]National Institute of Standards and Technology, Gaithersburg, Maryland 20899, U.S.A.

**Corresponding Author**

*E-mail : yongchen@purdue.edu ; park249@purdue.edu

**ABSTRACT**

We developed high-performance thermal interface materials (TIMs) based on few-layer graphene (FLG) composite, where FLG was prepared by the interlayer catalytic exfoliation (ICE) method. We experimentally demonstrated feasibility of FLG composites as TIMs by investigating their thermal and mechanical properties, and reliability. We measured the thermal interface resistance ($R_{int}$) between FLG composite TIMs (FLGTs) and copper and to be $3.2 \pm 1.7$ and $4.3 \pm 1.4$ mm$^2$K/W for 5 vol.% and 10 vol.% FLGTs at 330 K, respectively, comparable to or even lower than that of many commercial TIMs. In addition, the thermal conductivity ($\kappa_{TIM}$) of FLGTs is increased by an enhancement factor ($\beta$) of ~17 as the FLG concentration increases from 0 to 10 vol.%. We also characterized Vickers hardness and glass transition temperature ($T_g$) of our FLGTs. We find that our FLGTs are thermally and mechanically reliable within practical operating temperature and pressure ranges.

**INTRODUCTION**

A significant temperature discontinuity can occur at a thermal junction in a presence of a heat flux if the thermal junction has a high thermal interface resistance (a poor thermal link at the interface). This can cause serious problems in many applications such as industrial machinery, electronic, automobile, or medical devices as the generated heat flux has been significantly increasing due to the miniaturization of devices or high power used.[1–7] Hot spots on those devices or components with the poor thermal interface not cooled down efficiently thus, remaining at high temperatures, can deteriorate the performance, reliability, and life time of devices or components. The need to minimize the thermal interface resistance motivated the development of thermal interface materials (TIMs). To achieve an efficient heat conduction at thermal junctions, one can fill in surface irregularities (*e.g.* air gaps) at interfaces with TIMs, which are required to be thermally conductive and stable and have a low thermal interface resistance and a small (bond line) thickness.[8] In real applications, many types of TIMs are composite systems, consisting of matrices (*e.g.* silicone oil, hydrocarbon oil, and epoxy) and fillers (*e.g.* Ag, $Al_2O_3$, BN, carbon nanotube, and graphite).[8, 9] There have been a lot of efforts to investigate new types of fillers into composite TIMs with the aim to achieve better thermal and mechanical properties as well as a low production cost for thermal management applications.

As a promising candidate for a filler material into composite TIMs, graphene has received a lot of attention due to its excellent thermal and mechanical properties and relatively low cost.[10] However, in order to bring graphene-based composite TIMs into real applications, mass production of high-quality graphene is essential. Recently, X. Geng *et al*. developed the interlayer catalytic exfoliation (ICE) method that could enable massive production of high-quality FLG with a relatively large lateral size based on a simple and low-cost process,[11] compared with previous exfoliation approaches such as chemical reduction[12] and liquid-phase exfoliation.[13]

In this report, we experimentally demonstrate high-performance TIMs based on few-layer graphene (FLG) composite, where FLG has been prepared by ICE.[11] A number of recent studies on graphene (or graphite)-based composite TIMs have been reported.[14–18] However, most of these studies have focused on the measurement of thermal conductivity and its enhancement.[14–17] No direct measurement on the thermal interface resistance ($R_{int}$) has not been reported despite the promise of graphene-based composite as TIMs. Here, we present a first measurement of $R_{int}$ at the interface between FLG composite TIMs (FLGTs) and copper (Cu). We find that our measured $R_{int}$ (3~4 mm$^2$K/W) is comparable to or even lower than that of many commercial TIMs (*e.g.* $R_{int}$ of commercial Ag-filled adhesive TIMs with $R_{int}$ = ~10 mm$^2$K/W).[19, 20] We also characterize the thermal conductivity of FLGTs ($\kappa_{TIM}$) as well as thermal boundary resistance ($R_B$) between epoxy matrix and FLG inside the composite system, providing insights about enhancement of $\kappa_{TIM}$. In addition, we investigate the mechanical property and thermal stability of FLGTs that may give useful information related to their reliability.

**RESULT AND DISCUSSIONS**

FLG was prepared by the ICE method[11] and FLGTs were prepared by the simple process described in Figure 1a (see Experimental Methods). We centrifugally mixed FLG with epoxy and the composite mixture was cured in homemade metal molds. A representative Raman spectrum of a FLG flake prepared by the ICE technique is shown in Figure 1b. The defect-induced "D" peak intensity (normalized by graphene "G" peak intensity), I(D)/I(G), is measured to be less than 0.1, which indicates a low defect density and is consistent with the result in previous studies.[11] It is smaller by at least one order of magnitude than that of the chemically reduced graphene oxide.[21] It is almost comparable to graphene prepared by CVD[22] or LPE[13] as reported previously (more comprehensive characterizations of FLG prepared by the ICE method can be found in Ref.11). Figure 1c shows a SEM image of the cross-sectional structure at the interface of Cu-10 vol.% FLGT-Cu and it demonstrates a macroscopically intact bonding at the interface. Figure 1d shows a SEM image of crumpled and folded FLG flakes in our FLGT, similar to what has been seen in other graphene composites.[21, 23] We do not observe a structural orientation due to a random dispersion of FLG (see Figure S1 in Supporting Information for a similar observation in a few other samples of our FLGTs). In addition, we characterized the electrical conductivity of FLGTs, showing increased electrical conductivity with increasing filler volume fraction (see Supporting Information).

In order to evaluate the performance of FLGTs, we measured the thermal interface resistance ($R_{int}$) and the thermal conductivity ($\kappa_{TIM}$) of FLGTs using a modified ASTM 5470 method based on various earlier reports (see Experimental Methods and Supporting Information).[19, 24, 25] We measured spatial temperature profiles along the upper heat flux meter/Cu-FLGT-Cu/lower heat flux meter (where Cu-FLGT-Cu is a FLGT sandwiched by two copper (Cu) blocks, in direct contact with our flux meters that are also made of Cu, see Figure 2a inset). Representative profiles for 10 vol.% FLGTs with two different thicknesses at 330 K are shown in Figure 2a and b (the inset in Figure 2a depicts the measurement schematic). The upper heat flux ($Q_U$) and the lower heat flux ($Q_L$) can be extracted from the one-dimensional heat equation, $Q_U=\kappa_{Cu}(\Delta T_U/\Delta X_U)$ and $Q_L=\kappa_{Cu}(\Delta T_L/\Delta X_L)$, where $\kappa_{Cu}$ is the reference thermal conductivity of oxygen free copper,[26] $\Delta T_U$=T1-T2, $\Delta T_L$=T5-T6, and $\Delta X_U=\Delta X_L$=20 mm (distance between RTDs (resistance temperature detectors, Lakeshore Pt-103) 1-2 and between 5-6). Then we calculated the average heat flux, $Q_{AV}=(Q_U+Q_L)/2$. The total thermal resistance ($R=\Delta T/Q_{AV}$, where $\Delta T$=T3-T4) of the FLGT and portions of the two copper blocks (between T3 and T4) can be written as $R=R_{Cu1}+R_{TIM}+R_{Cu2}$ and $R_{TIM}=2R_{int}+t/\kappa_{TIM}$, where $R_{Cu1}$ and $R_{Cu2}$ are thermal resistances due to corresponding portions of the upper and lower copper blocks (that can be calculated from $\kappa_{Cu}$ and the distances from RTDs 3 and 4 to upper and lower thermal interfaces, respectively), $R_{int}$ is the thermal interface resistance at the FLGT-Cu interface, t is the thickness of the FLGT, and $\kappa_{TIM}$ is the thermal conductivity of the FLGT. We measured $R_{int}$ and $\kappa_{TIM}$ of FLGTs from the thickness-dependent $R_{TIM}$ curve ($R_{TIM}$ vs. t), using a linear fit as shown in Figure 2c. The slope of the linear fit and the intercept at y-axis are equal to $1/\kappa_{TIM}$ and $2R_{int}$, respectively. At 330 K, we measured $R_{int}$ of 5 vol.% and 10 vol.% FLGTs to be 3.2 ± 1.7 and 4.3 ± 1.4 mm²K/W, and $\kappa_{TIM}$ of 5 vol.% and 10 vol.% FLGTs to be and 2.8 ± 0.2 and 3.9 ± 0.3 W/mK, respectively. Furthermore, we observe that $R_{int}$ and $\kappa_{TIM}$ for 5 vol.% and 10 vol.% FLGTs do not show appreciable dependence on temperature from 300 to 370 K (see Figure S3 in Supporting Information). The performance of FLGTs is comparable to many commercial TIMs, carbon nanotube arrays, and vertically aligned multi-layer graphene coated with indium as shown in Table 1.

Figure 2d shows $\kappa_{TIM}$ of FLGTs as a function of FLG concentration (volume fraction f). We find that $\kappa_{TIM}$ of FLGTs increases from 0.21 ± 0.03 to 3.87 ± 0.28 W/mK at room temperature when the FLG concentration increases from 0 vol.% to 10 vol.% (enhancement factor (β) of ~17 at 10 vol.% in the inset of Figure 2d). However, we observe that a noticeable sublinear behavior of $\kappa_{TIM}$ versus filler (FLG) concentration in contrast to the previous result for graphene/multi-layer graphene (GMLG) epoxy composites, which shows an almost linear enhancement when adding GMLG from 0 to 10 vol.%.[15] We used a modified effective medium approximation (EMA) theory[33, 34] to analyze the thermal boundary resistance ($R_B$) between epoxy and FLG in FLGTs. We can write $\kappa_{TIM}$ of FLGTs as (a detail derivation can be found in Supporting Information)

$$\kappa_{TIM} = \frac{3\kappa_m + 2f\left(\frac{\kappa_{px}}{1 + \frac{2R_B\kappa_{px}}{L}} - \kappa_m\right)}{3 - f\left(1 - \frac{2R_B\kappa_m}{h}\right)}, \quad (1)$$

where $\kappa_m$ is the thermal conductivity of epoxy matrix (0.21 W/mK), f is the volume fraction of FLG, $\kappa_{px}$ is the in-plane thermal conductivity of FLG (~1670 W/mK),[35–37] h is the typical thickness of FLG (~2 nm),[11] and L is the typical lateral size of FLG (~10 μm).[11] We find that $R_B$ between epoxy and FLG, by fitting the data in Figure 2d to Eq. (1), is ~6×10$^{-8}$ m$^2$W/K. The estimated $R_B$ is higher than that of GMLG composites (3.5×10$^{-9}$ m$^2$W/K) by a factor of ~17, which leads to the relatively low enhancement in $\kappa_{TIM}$ at 10 vol.%, compared with that in Ref.15. On the other hand, $R_B$ is still lower than that of untreated graphite nanoplatelet composites by a factor of ~11.[38] We note that $R_B$ at the interface of graphene and dissimilar materials can be as low as ~10$^{-9}$ m$^2$W/K based on earlier theoretical and experimental studies (*e.g.* $R_B$ of graphene-octane, graphene-copper, or graphene-SiO$_2$).[39–41] We speculate that our relatively large $R_B$ at FLG-epoxy interface possibly originates from structural complexities of FLG (*e.g.* edge and surface roughness, or functional groups induced by the ICE process). However, we expect that $\kappa_{TIM}$ of FLGTs can be enhanced more with chemical modification of FLG or with a use of hybrid fillers that may lower $R_B$.[42, 43]

We further investigated the mechanical property of FLGTs using a Vickers hardness test (see Figure S4 in Supporting Information) as shown in Figure 3a. The hardness of FLGTs is increased by ~17 % and reaches a peak value of 242 MPa when the FLG concentration increases from 0 (pure epoxy) to 0.25 vol.% and it gradually decreases above 0.25 vol.%. The hardness of FLGTs becomes lower than that of pure epoxy above 1 vol.%. A recent study has shown a similar trend in graphene nanoplatelet composites (where the highest hardness (~220 MPa) is found at ~0.2 vol.% and it starts to decrease beyond that volume fraction).[44] The increase of hardness at the relatively low concentration regime (< 0.5 vol.%) indicates an effective load transfer from epoxy matrix to FLG, whereas at the high concentration regime, non-uniform dispersion of fillers such as agglomerations can impede the load transfer and distribution through composites, causing the decrease of hardness.[44, 45]

TIMs are often exposed to high temperatures so that thermal stability is an important factor in real applications of TIMs. In order to investigate the thermal stability of FLGTs, we measured the glass transition temperature ($T_g$) of FLGTs with varying the FLG concentration as shown in Figure 3b, using a differential scanning calorimetry (DSC) method (see Figure S5 in Supporting Information). In general, $T_g$ marks a phase transition from a glassy state ($T < T_g$) to a rubbery state ($T > T_g$) in polymers, where the mechanical and thermal properties are often degraded above $T_g$. We observe that $T_g$ of FLGTs increases with the increasing FLG concentration, suggesting improvement in the thermal stability of FLGTs. We find that $T_g$ is enhanced by ~50 K when increasing the FLG concentration from 0 to 10 vol.%. We note that most of the $T_g$ enhancement occurs at the low concentration regime from 0 to 1 vol.%. Similar and notable enhancement at low filler concentration was also reported in well-dispersed functionalized graphene/PMMA composite and it was explained by a dispersion state transition between a discrete interphase region and a percolated interphase region, leading to a change in mobility of the matrix polymer.[46] In general, well-dispersed nanofillers in polymer composites can efficiently confine the motion of polymer chains, which may increase $T_g$.[47, 48] In addition, in our FLGTs, a relatively low $T_g$ has been observed as compared with that in previous studies based on graphite or graphene fillers.[42, 49] It may be due to the low curing temperature (120 °C) used in our sample preparation, possibly causing incomplete curing (a similar result is shown in Ref.50). We expect that an optimized curing process can further improve the thermal stability.

It is also important to note that TIMs often undergo thermal and mechanical stresses (*e.g.* multiple thermal cycles). Such environments can deteriorate the performance of TIMs in a long-term use so that relevant tests are important. We measured $R_{TIM}$ of representative 49 μm-thick 5 vol.% and 45 μm-thick 10 vol.% FLGTs with increasing pressure from 0.14 MPa to 1 MPa as shown in Figure 4a. We do not observe any appreciable change of $R_{TIM}$ for each of samples under different pressures, indicating that the thermal performance is not degraded within the measurement range (a realistic pressure range in CPU packaging is ~0.12 MPa[19]).

For further investigation of the reliability under thermal cycling, we monitored change in $R_{TIM}$ of representative FLGTs as shown in Figure 4b when repeatedly changing the average temperature of FLGTs between 300 and 370 K (lower than $T_g$ of FLGTs). No noticeable change in $R_{TIM}$ for each of FLGTs is observed after the ten thermal cycles tested. It suggests, for example, that there is no significant thermal cycling induced delamination (which would increase $R_{int}$) during this test.

**CONCLUSION**

In summary, we investigated thermal properties of FLGTs ($R_{int}$ and $\kappa_{TIM}$) to demonstrate feasibility of FLG composites as TIMs. We measured $R_{int}$ to be 3.2 ± 1.7 and 4.3 ± 1.4 mm$^2$K/W for 5 vol.% and 10 vol.% FLGTs at 330 K, respectively. The measured $R_{int}$ is comparable to many commercial TIMs and we expect that there is further room for improving $R_{int}$. Recent studies imply that $R_{int}$ can be further improved by enhancing the real bonding area with tuning the wetting parameter of TIMs (*e.g.* surface energy and contact angle).[51, 52] It suggests that future work can focus more on optimizing the real bonding area by investigating the wettability change of FLGTs during the assembling/curing process, rather than just focusing on studying methods to improve the thermal conductivity of bulk TIMs. We also find that $\kappa_{TIM}$ of FLGTs is improved by an enhancement factor β of ~17 with increasing the FLG concentration from 0 to 10 vol.%. The highest $\kappa_{TIM}$ is measured to 3.87 ± 0.28 W/mK at 10 vol.% at room temperature. We observe a sublinear behavior of $\kappa_{TIM}$ versus FLG concentration with the relatively large fitted $R_B$ at the FLG-epoxy matrix interface. In addition, we find that the thermal stability as characterized by $T_g$ of FLGTs is enhanced with adding FLG and that FLGTs are not vulnerable to the thermal stress during multiple thermal cycles tested.

**EXPERIMENTAL METHODS**

*Preparation of FLG* : We preapred FLG based on the previous report.[11] High-quality FLG powder was prepared by interlayer catalytic exfoliation (ICE) of ferric chloride ($FeCl_3$)-graphite intercalation compound. The $FeCl_3$-intercalated graphite (FIG) compound was synthesized by a conventional two-zone vapour transport technique. As-synthesized FIG and $H_2O_2$ (30%) were loaded into a reactive bottle at room temperature for about 2 hours. The FIG was exfoliated into long worm-like graphite consisting of interconnecting graphene layers, and then a gentle and short time (5 minutes) ultra-sonication was performed to obtain FLG.

*Preparation of FLG composite TIMs* : For the FLGT formulation, we assumed the density of graphene and epoxy to be ~2.2 and ~1.2 $g/cm^3$, repectively, and maintained epoxy resin (Epon 862, Miller-Stephenson)/curing agent (Epikure W, Miller-Stephenson) ratio = 100/26.4 by weight. We dried as-preapred FLG for 1 day at 150 °C in order to remove residual solvents. Then, we added FLG to Epon 862 based on the vol.% caculation and blended this mixture with acetone. The dispersion solution was ultra-sonicated for 10 minutes and dired in ambient condition overnight. It was additionally dried at 80 °C in a vacuum oven until acetone is fully removed. We centrifugally mixed pre-weighted Epikure W with the mixture of FLG/Epon 862 at 2000 rpm for 30 minutes using Thinky ARE-310. The uncured composite was poured or pasted into metal molds and it was cured at 100 °C for 2 hours followed by additional post-curing at 120 °C for 4 hours (see Figure 1a and S2b). In order to prepare the FLGT sandwiched by copper blocks (Cu-FLGT-Cu), we pasted the uncured composite mixture on the surface of copper blocks (each one is 10 mm-long with diameter of 19.05 mm and we drilled a hole at the vertical center of each copper block for inserting the RTD as shown in Figure 2a and S2), and assembled them together. We controlled the thickness of FLGTs (t) by changing the thickness of G-10 (flame-retardant garolite) shims (inserted between two Cu blocks, where the area of each shim is 1~2mm$^2$, less than ~3 % of total area of the copper block surface, leading negligible effect on the thermal transport). After the assembling process, the sandwiched structure (Cu-FLGT-Cu) was cured all toghether by following the same curing process described above (see Figure 1a and S2a).

*Raman and SEM characterization* : We performed Raman spectroscopy on a FLG flake using a Horiba Jobin Yvon Xplora confocal Raman microscope with a 532 nm laser (power=~1.4 mW) and a 100x objective. Cross-sectional structures of FLGTs were studied by a scanning electron microscope (SEM) (Hitachi S-4800). We milled the cured Cu-FLGTs-Cu sample and polished the cross-sectional surface using alumina nanoparticles (50 nm) in order to investigate the interface of the Cu-FLGT-Cu.

*Thermal conductivity ($\kappa_{TIM}$) and thermal interfac resistance ($R_{int}$) measurements* : We developed a modified ASTM D5470 system based on the previous report.[19, 24, 25] Two oxygen-free copper (OFC) rods (diameter=19.05 mm) with RTDs were used as the heat flux meters. The FLGT sample was inserted between the upper and lower heat flux meters described in Supporting Information. We used a cartridge heater, which is located on the top of the upper heat flux meter (Figure 2a inset) and controlled by a temperature controller (Lakeshore 340), to increase the average sample temperature, and kept supplying water to the cold plate for cooling. The pressure was controlled by four spring-loaded clamps. The temperature profile (temperature readings of all RTDs) was recorded when the entire system reached a steady state condition and all measurements were conducted at a vacuum condition (<~20mTorr) with 0.14 MPa of pressure applied unless otherwise noted. For the thermal cycling test, we increased the sample temperature from 300 K to 370 K with a ramping rate of ~10K/minute after the initial characterizations at room temperature, and then the sample was naturally cooled down to 300 K again. The temperature profile was recorded only after a steady state was reached and we repeated those procedures for multiple cycles. During all measurements, we observed a negligible heat flux difference between the upper and lower heat flux meters, indicating the measured heat flux is reliable without a significant heat loss or generation.

*Glass transition temperature ($T_g$) measurement* : We measured $T_g$ using Jade DSC. FGLTs were cut into thin disks (~10 mg) and were mounted in aluminum pans. FLGTs were heated with a ramping rate of 10 K/min under $N_2$ condition with a flow rate of 20 mL/minute. We calculated $T_g$ based on a half-step height method.

*Vickers hardness test* : We performed the Vickers hardness test using LECO LV-100. FLGTs with thickness ~1 mm were prepared for the measurement. The loading force was 1 kgf (= 9.8N) and the dwell time was 15 seconds. The FLGT surface after loading was analyzed by an optical microscope (Olympus BX51M) (see Figure S4 in Supporting Information).

*Electrical conductivity measurement* : FLGTs were cut into square slabs and we deposited Cr/Au(20/180nm) as electrodes on the each surface of FLGTs using e-beam evaporation. The channel area and the channel length were 16-25mm$^2$ and 0.4-0.5 mm, respectively. We measured the electrical conductivity (σ) of FLGTs using a source-meter (Keithley 2400) and a multimeter (HP 34401A) based on a four-terminal method.

ACKNOWLEDGEMENT

We thank Dr. Rebecca Kramer at Purdue University for access to the centrifugal mixer and thank Mr. Jongbeom Kim and Nirajan Mandal at Purdue University for helping us to do structural characterizations. This work was partly supported by the Purdue Cooling Technologies Research Center (CTRC), a National Science Foundation (NSF) industry/university cooperative research.

REFERENCES

(1) Moore, A. L.; Shi, L. Emerging Challenges and Materials for Thermal Management of Electronics. *Mater. Today* **2014**, *17*, 163–174.

(2) Pop, E.; Sinha, S.; Goodson, K. E. Heat Generation and Transport in Nanometer-Scale Transistors. *Proc. IEEE* **2006**, *94*, 1587–1601.

(3) Schelling, P. K.; Shi, L.; Goodson, K. E. Managing Heat for Electronics. *Mater. Today* **2005**, *8*, 30–35.

(4) Lazzi, G. Thermal Effects of Bioimplants. *Eng. Med. Biol. Mag. IEEE* **2005**, *24*, 75–81.

(5) Mallik, S.; Ekere, N.; Best, C.; Bhatti, R. Investigation of Thermal Management Materials for Automotive Electronic Control Units. *Appl. Therm. Eng.* **2011**, *31*, 355–362.

(6) Weng, C.-J. Advanced Thermal Enhancement and Management of LED Packages. *Int. Commun. Heat Mass Transf.* **2009**, *36*, 245–248.

(7) Duan, X.; Naterer, G. F. Heat Transfer in Phase Change Materials for Thermal Management of Electric Vehicle Battery Modules. *Int. J. Heat Mass Transf.* **2010**, *53*, 5176–5182.

(8)  Prasher, R. Thermal Interface Materials: Historical Perspective, Status, and Future Directions. *Proc. IEEE* **2006**, *94*, 1571–1586.

(9)  Chung, D. D. L. Thermal Interface Materials. *J. Mater. Eng. Perform.* **2001**, *10*, 56–59.

(10) Geim, A. K. Graphene: Status and Prospects. *Science* **2009**, *324*, 1530–1534.

(11) Geng, X.; Guo, Y.; Li, D.; Li, W.; Zhu, C.; Wei, X.; Chen, M.; Gao, S.; Qiu, S.; Gong, Y.; *et al.* Interlayer Catalytic Exfoliation Realizing Scalable Production of Large-Size Pristine Few-Layer Graphene. *Sci. Rep.* **2013**, *3*, 1134.

(12) Stankovich, S.; Dikin, D. A.; Piner, R. D.; Kohlhaas, K. A.; Kleinhammes, A.; Jia, Y.; Wu, Y.; Nguyen, S. T.; Ruoff, R. S. Synthesis of Graphene-Based Nanosheets via Chemical Reduction of Exfoliated Graphite Oxide. *Carbon* **2007**, *45*, 1558–1565.

(13) Hernandez, Y.; Nicolosi, V.; Lotya, M.; Blighe, F. M.; Sun, Z.; De, S.; McGovern, I. T.; Holland, B.; Byrne, M.; Gun'Ko, Y. K.; *et al.* High-Yield Production of Graphene by Liquid-Phase Exfoliation of Graphite. *Nat. Nano.* **2008**, *3*, 563–568.

(14) Yu, A.; Ramesh, P.; Itkis, M. E.; Bekyarova, E.; Haddon, R. C. Graphite Nanoplatelet−Epoxy Composite Thermal Interface Materials. *J. Phys. Chem. C* **2007**, *111*, 7565–7569.

(15) Shahil, K. M. F.; Balandin, A. A. Graphene–Multilayer Graphene Nanocomposites as Highly Efficient Thermal Interface Materials. *Nano Lett.* **2012**, *12*, 861–867.

(16) Tian, X.; Itkis, M. E.; Bekyarova, E. B.; Haddon, R. C. Anisotropic Thermal and Electrical Properties of Thin Thermal Interface Layers of Graphite Nanoplatelet-Based Composites. *Sci. Rep.* **2013**, *3*, 1710.

(17) Tang, B.; Hu, G.; Gao, H.; Hai, L. Application of Graphene as Filler to Improve Thermal Transport Property of Epoxy Resin for Thermal Interface Materials. *Int. J. Heat Mass Transf.* **2015**, *85*, 420–429.

(18) Lin, C.; Chung, D. D. L. Graphite Nanoplatelet Pastes vs. Carbon Black Pastes as Thermal Interface Materials. *Carbon* **2009**, *47*, 295–305.

(19) Gwinn, J. P.; Webb, R. L. Performance and Testing of Thermal Interface Materials. *Microelectronics J.* **2003**, *34*, 215–222.

(20) Teertstra, P. Thermal Conductivity and Contact Resistance Measurements for Adhesives. In *ASME 2007 InterPACK Conference collocated with the ASME/JSME 2007 Thermal Engineering Heat Transfer Summer Conference*; pp 381–388.

(21) Park, W.; Hu, J.; Jauregui, L. A.; Ruan, X.; Chen, Y. P. Electrical and Thermal Conductivities of Reduced Graphene Oxide/polystyrene Composites. *Appl. Phys. Lett.* **2014**, *104*, 113101.

(22) Li, X.; Cai, W.; An, J.; Kim, S.; Nah, J.; Yang, D.; Piner, R.; Velamakanni, A.; Jung, I.; Tutuc, E.; *et al.* Large-Area Synthesis of High-Quality and Uniform Graphene Films on Copper Foils. *Science* **2009**, *324*, 1312–1314.

(23) Stankovich, S.; Dikin, D. A.; Dommett, G. H. B.; Kohlhaas, K. M.; Zimney, E. J.; Stach, E. A.; Piner, R. D.; Nguyen, S. T.; Ruoff, R. S. Graphene-Based Composite Materials. *Nature* **2006**, *442*, 282–286.

(24) Gwinn, J. P.; Saini, M.; Webb, R. L. Apparatus for Accurate Measurement of Interface Resistance of High Performance Thermal Interface Materials. In *Thermal and Thermomechanical Phenomena in Electronic Systems, 2002. ITHERM 2002. The Eighth Intersociety Conference on*; pp 644–650.

(25) Ngo, Q.; Cruden, B. A.; Cassell, A. M.; Sims, G.; Meyyappan, M.; Li, J.; Yang, C. Y. Thermal Interface Properties of Cu-Filled Vertically Aligned Carbon Nanofiber Arrays. *Nano Lett.* **2004**, *4*, 2403–2407.

(26) Hong, S.-T.; Herling, D. R. Effects of Surface Area Density of Aluminum Foams on Thermal Conductivity of Aluminum Foam-Phase Change Material Composites. *Adv. Eng. Mater.* **2007**, *9*, 554–557.

(27) Leong, C. K.; Chung, D. D. L. Carbon Black Dispersions and Carbon-Silver Combinations as Thermal Pastes That Surpass Commercial Silver and Ceramic Pastes in Providing High Thermal Contact Conductance. *Carbon* **2004**, *42*, 2323–2327.

(28) Howe, T. A.; Leong, C.-K.; Chung, D. D. L. Comparative Evaluation of Thermal Interface Materials for Improving the Thermal Contact between an Operating Computer Microprocessor and Its Heat Sink. *J. Electron. Mater.* **2006**, *35*, 1628–1635.

(29) Xu, J.; Fisher, T. S. Enhancement of Thermal Interface Materials with Carbon Nanotube Arrays. *Int. J. Heat Mass Transf.* **2006**, *49*, 1658–1666.

(30) Gao, Z. L.; Zhang, K.; Yuen, M. M. F. Fabrication of Carbon Nanotube Thermal Interface Material on Aluminum Alloy Substrates with Low Pressure CVD. *Nanotechnology* **2011**, *22*, 265611.

(31) Zhang, K.; Chai, Y.; Yuen, M. M. F.; Xiao, D. G. W.; Chan, P. C. H. Carbon Nanotube Thermal Interface Material for High-Brightness Light-Emitting-Diode Cooling. *Nanotechnology* **2008**, *19*, 215706.

(32) Liang, Q.; Yao, X.; Wang, W.; Liu, Y.; Wong, C. P. A Three-Dimensional Vertically Aligned Functionalized Multilayer Graphene Architecture: An Approach for Graphene-Based Thermal Interfacial Materials. *ACS Nano* **2011**, *5*, 2392–2401.

(33) Nan, C.-W.; Birringer, R.; Clarke, D. R.; Gleiter, H. Effective Thermal Conductivity of Particulate Composites with Interfacial Thermal Resistance. *J. Appl. Phys.* **1997**, *81*, 6692–6699.

(34) Nan, C.-W.; Liu, G.; Lin, Y.; Li, M. Interface Effect on Thermal Conductivity of Carbon Nanotube Composites. *Appl. Phys. Lett.* **2004**, *85*, 3549–3551.

(35) Sadeghi, M. M.; Jo, I.; Shi, L. Phonon-Interface Scattering in Multilayer Graphene on an Amorphous Support. *Proc. Natl. Acad. Sci. U. S. A.* **2013**, *110*, 16321–16326.

(36) Ghosh, S.; Bao, W.; Nika, D. L.; Subrina, S.; Pokatilov, E. P.; Lau, C. N.; Balandin, A. A. Dimensional Crossover of Thermal Transport in Few-Layer Graphene. *Nat. Mater.* **2010**, *9*, 555–558.

(37) Chu, K.; Li, W.; Tang, F. Flatness-Dependent Thermal Conductivity of Graphene-Based Composites. *Phys. Lett. A* **2013**, *377*, 910–914.

(38) Hung, M.; Choi, O.; Ju, Y. S.; Hahn, H. T. Heat Conduction in Graphite-Nanoplatelet-Reinforced Polymer Nanocomposites. *Appl. Phys. Lett.* **2006**, *89*, 023117.

(39) Konatham, D.; Striolo, A. Thermal Boundary Resistance at the Graphene-Oil Interface. *Appl. Phys. Lett.* **2009**, *95*, 163105.

(40) Chang, S.-W.; Nair, A. K.; Buehler, M. J. Geometry and Temperature Effects of the Interfacial Thermal Conductance in Copper– and Nickel–graphene Nanocomposites. *J. Phys. Condens. Matter* **2012**, *24*, 245301.

(41) Chen, Z.; Jang, W.; Bao, W.; Lau, C. N.; Dames, C. Thermal Contact Resistance between Graphene and Silicon Dioxide. *Appl. Phys. Lett.* **2009**, *95*, 161910.

(42) Ganguli, S.; Roy, A. K.; Anderson, D. P. Improved Thermal Conductivity for Chemically Functionalized Exfoliated Graphite/epoxy Composites. *Carbon* **2008**, *46*, 806–817.

(43) Chen, L.; Sun, Y.-Y.; Lin, J.; Du, X.-Z.; Wei, G.-S.; He, S.-J.; Nazarenko, S. Modeling and Analysis of Synergistic Effect in Thermal Conductivity Enhancement of Polymer Composites with Hybrid Filler. *Int. J. Heat Mass Transf.* **2015**, *81*, 457–464.

(44) Wang, Y.; Yu, J.; Dai, W.; Song, Y.; Wang, D.; Zeng, L.; Jiang, N. Enhanced Thermal and Electrical Properties of Epoxy Composites Reinforced with Graphene Nanoplatelets. *Polym. Compos.* **2015**, *36*, 556–565.

(45) Chatterjee, S.; Wang, J. W.; Kuo, W. S.; Tai, N. H.; Salzmann, C.; Li, W. L.; Hollertz, R.; Nüesch, F. A.; Chu, B. T. T. Mechanical Reinforcement and Thermal Conductivity in Expanded Graphene Nanoplatelets Reinforced Epoxy Composites. *Chem. Phys. Lett.* **2012**, *531*, 6–10.

(46) Ramanathan, T.; Abdala, A. A.; Stankovich, S.; Dikin, D. A.; Herrera-Alonso, M.; Piner, R. D.; Adamson, D. H.; Schniepp, H. C.; Chen, X.; Ruoff, R. S.; *et al.* Functionalized Graphene Sheets for Polymer Nanocomposites. *Nat. Nanotechnol.* **2008**, *3*, 327–331.

(47) Bansal, A.; Yang, H.; Li, C.; Cho, K.; Benicewicz, B. C.; Kumar, S. K.; Schadler, L. S. Quantitative Equivalence between Polymer Nanocomposites and Thin Polymer Films. *Nat. Mater.* **2005**, *4*, 693–698.

(48) Rittigstein, P.; Priestley, R. D.; Broadbelt, L. J.; Torkelson, J. M. Model Polymer Nanocomposites Provide an Understanding of Confinement Effects in Real Nanocomposites. *Nat. Mater.* **2007**, *6*, 278–282.

(49) Naebe, M.; Wang, J.; Amini, A.; Khayyam, H.; Hameed, N.; Li, L. H.; Chen, Y.; Fox, B. Mechanical Property and Structure of Covalent Functionalised Graphene/epoxy Nanocomposites. *Sci. Rep.* **2014**, *4*, 4375.

(50) Gu, H.; Tadakamalla, S.; Zhang, X.; Huang, Y.; Jiang, Y.; Colorado, H. A.; Luo, Z.; Wei, S.; Guo, Z. Epoxy Resin Nanosuspensions and Reinforced Nanocomposites from Polyaniline Stabilized Multi-Walled Carbon Nanotubes. *J. Mater. Chem. C* **2013**, *1*, 729–743.

(51) Prasher, R. S. Surface Chemistry and Characteristics Based Model for the Thermal Contact Resistance of Fluidic Interstitial Thermal Interface Materials. *J. Heat Transfer* **2001**, *123*, 969–975.

(52) Somé, S. C.; Delaunay, D.; Faraj, J.; Bailleul, J.-L.; Boyard, N.; Quilliet, S. Modeling of the Thermal Contact Resistance Time Evolution at Polymer–mold Interface during Injection Molding: Effect of Polymers' Solidification. *Appl. Therm. Eng.* **2015**, *84*, 150–157.

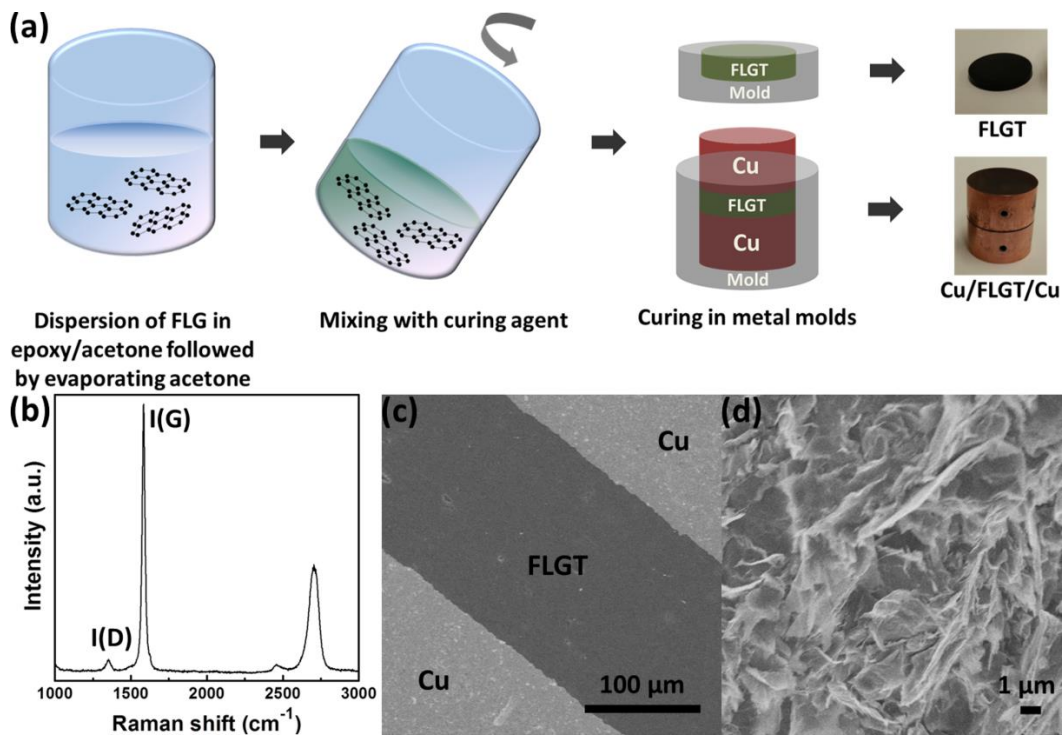

**Figure 1**. (a) Schematic illustration of the FLGT and Cu/FLGT/Cu preparation procedure. (b) Raman spectrum of a FLG flake prepared by the ICE method. (c) SEM image of the interface at Cu-10 vol.% FLGT-Cu (at ×300k magnification). (d) SEM image of 10 vol.% FLGT (at ×5k magnification).

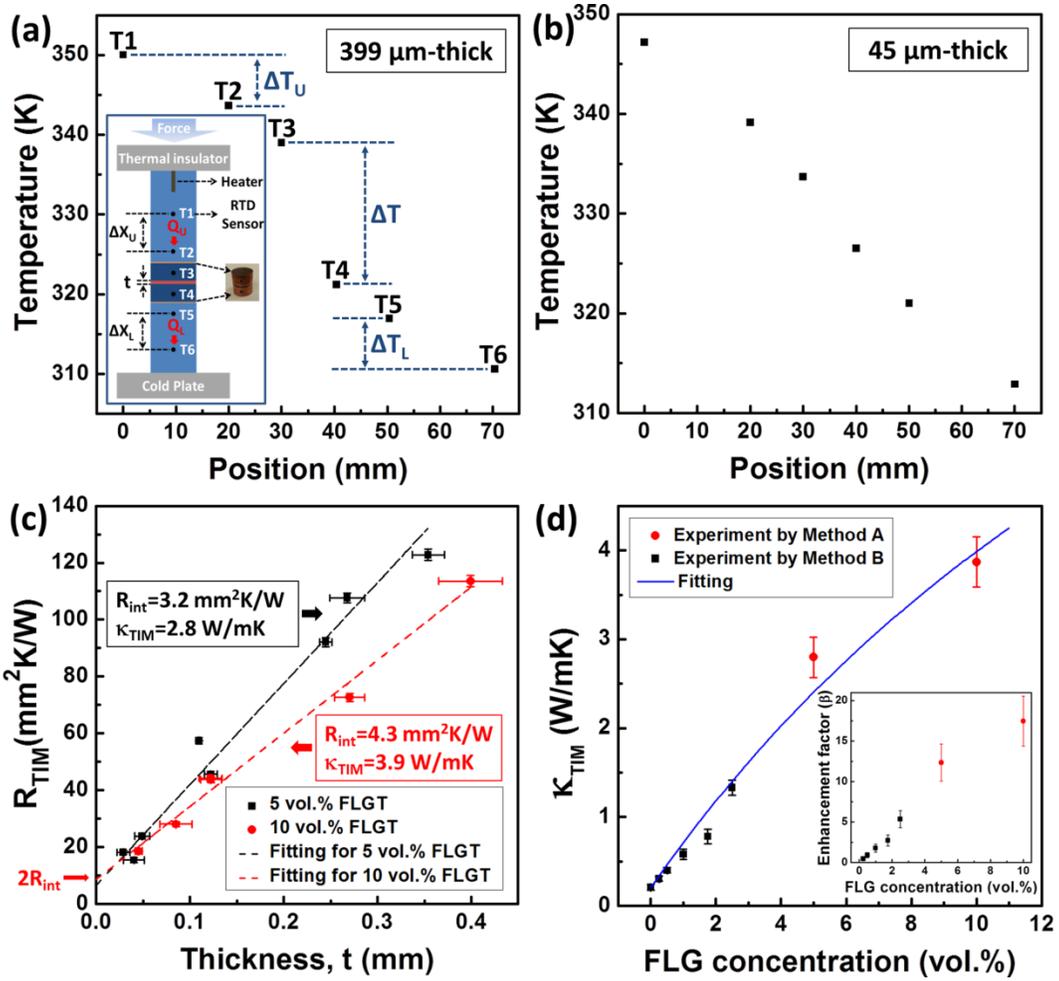

**Figure 2**. (a,b) Temperature profiles along the upper heat flux meter/Cu-FLGT-Cu/lower heat flux meter for (a) 399 μm-thick and (b) 45 μm-thick 10 vol.% FLGTs at 330 K (inset in (a), illustration of the heat flux meters where $Q_U$ and $Q_L$ represent the heat flux along the upper and lower heat flux meters, respectively, t is the thickness of the FLGT, $\Delta T_U$=T1-T2, $\Delta T_L$=T5-T6, and $\Delta X_U$=$\Delta X_L$=20 mm (distance between RTD sensors)). (c) Thickness-dependent $R_{TIM}$ of 5 vol.% and 10 vol.% FLGTs (intercepts of the dashed linear fit represent $2R_{int}$). (d) $\kappa_{TIM}$ of FLGTs as a function of FLG concentration (inset in (d), enhancement factor ($\beta$=($\kappa_{TIM}$−$\kappa_m$)/$\kappa_m$, where $\kappa_m$ is the thermal conductivity of pure epoxy) of FLGTs as a function of FLG concentration). We measured $\kappa_{TIM}$ of 5 and 10 vol.% FLGTs by Method A and measured $\kappa_{TIM}$ of 0-2.5 vol.% FLGTs by Method B (see Figure S2 in Supporting Information).

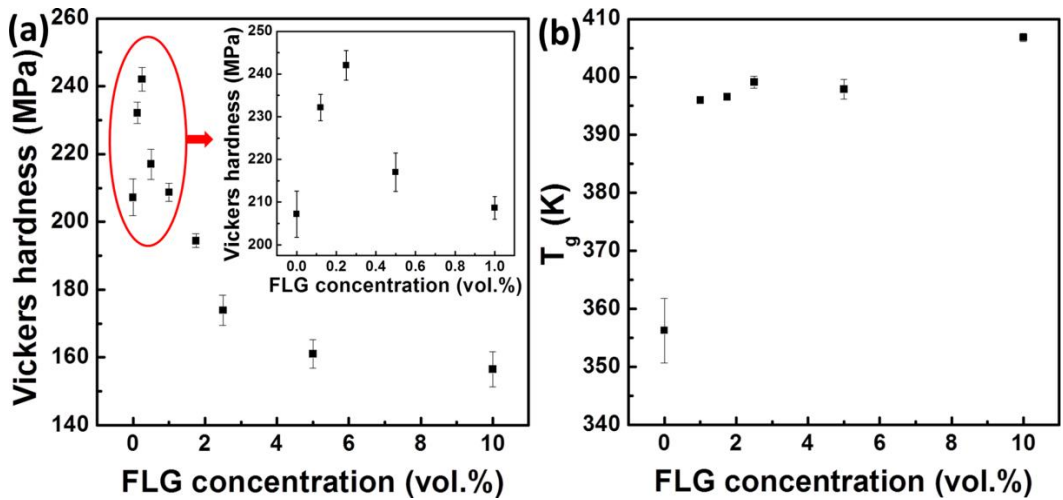

**Figure 3**. (a) Vickers hardness of FLGTs as a function of FLG concentration (inset, Vickers hardness from 0 to 1 vol.% FLGTs). (b) Glass transition temperature ($T_g$) of FLGTs as a function of FLG concentration.

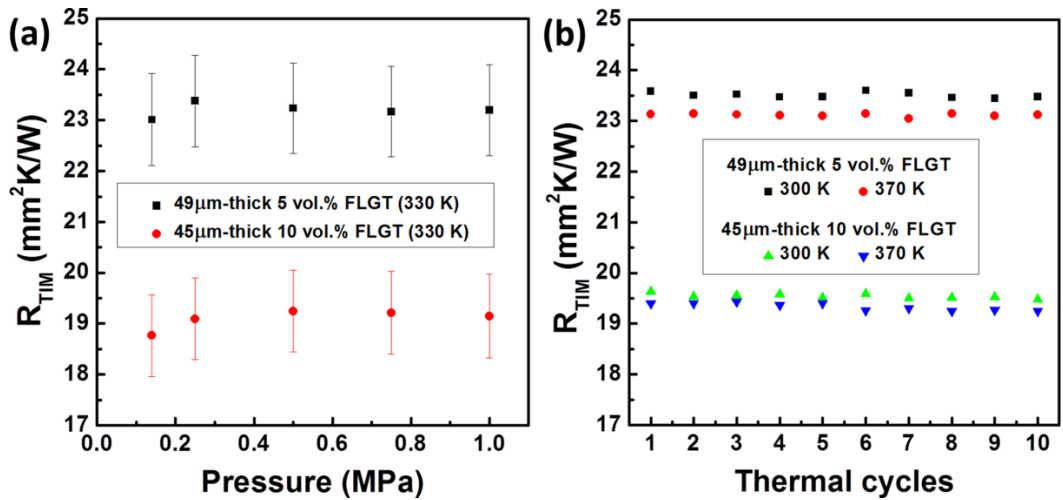

**Figure 4**. (a) $R_{TIM}$ of 49 μm-thick 5 vol.% and 45 μm-thick 10 vol.% FLGTs at 330 K under various pressures. (b) $R_{TIM}$ of 49 μm-thick 5 vol.% and 45 μm-thick 10 vol.% FLGTs under multiple thermal cycles.

**Table 1**. Comparison of $R_{int}$ and $\kappa_{TIM}$ of commercial TIMs and various carbon-based TIMs including our FLGTs

| Type of TIMs | $\kappa_{TIM}$ (W/mK) | $R_{int}$ (mm$^2$K/W) |
|---|---|---|
| RTV silicone[20] | 0.53 | 7.9 |
| Al-filled epoxy putty[20] | 0.65 | 10.3 |
| Al-filled 2-part epoxy bonding resin[20] | 0.84 | 31 |
| Silver-filled thermoplastic[20] | 7.8 | 10.3 |
| Silver-filled grease[27,28, a] | 8.4-9 | 5.6-9.2 |
| Carbon nanotube (CNT) array[29, a] | NA | 19.8 |
| CNT array[30, a] | NA | 14.6 |
| CNT array[31, a] | NA | 7 |
| Multilayer graphene with indium melt[32] | 75.5 | 5.1 |
| Graphite nanoplatelet/ polyol-ester oil[18] | 0.48 | 3.1 |
| 5 vol.% FLGT | 2.8 | 3.2 |
| 10 vol.% FLGT | 3.9 | 4.3 |

[a] $R_{TIM}$ is reported instead of $R_{int}$ but the bulk thermal resistance is small due to a large $\kappa_{TIM}$ and a small bond line thickness.

# Supporting Information

**1. Supplementary figures**

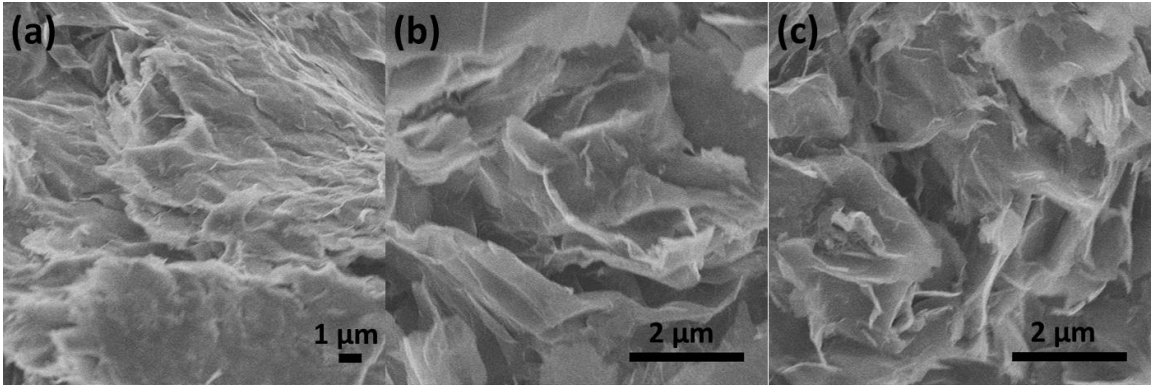

Figure S1. (a-c) SEM images of (a) 5 vol. % FLGT (at ×5k magnification), (b) 5 vol. % FLGT (at ×13k magnification), and (c) 10 vol.% FLGT (at ×13k magnification).

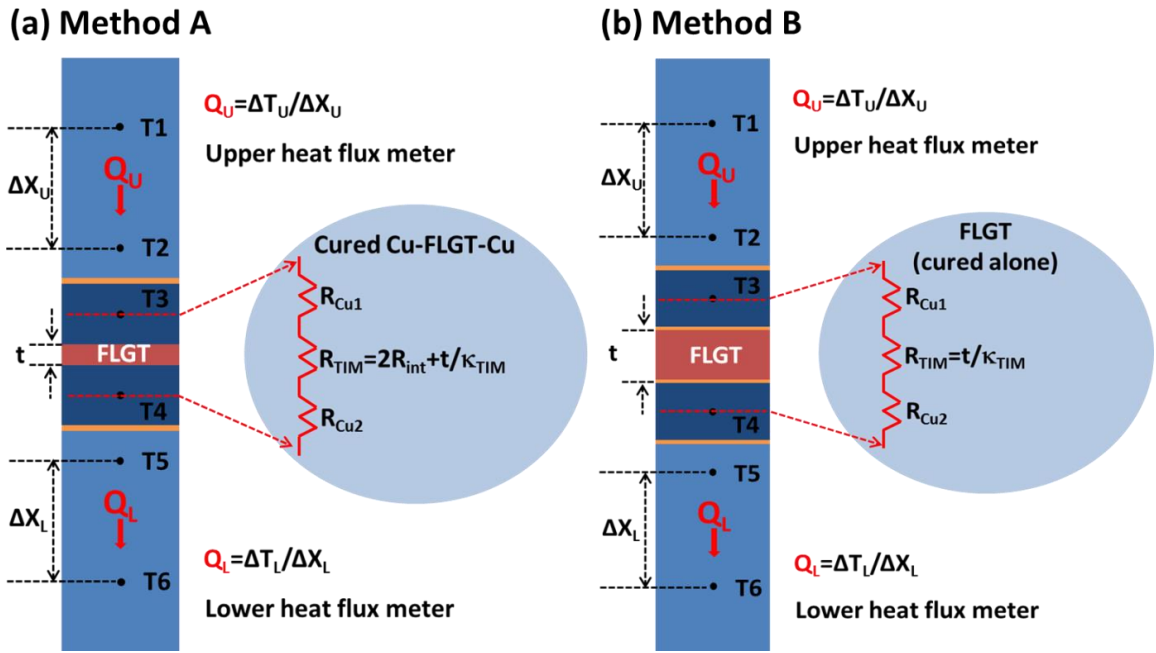

Figure S2. (a,b) Illustration of sample assemblies for (a) $\kappa_{TIM}$ and $R_{int}$ measurements on FLGTs by Method A and (b) $\kappa_{TIM}$ measurement on FLGTs by Method B (interfaces are exaggerated and orange color indicates the thermal grease applied).

For the $\kappa_{TIM}$ and $R_{int}$ measurements on 5 and 10 vol.% FLGTs as shown in Figure S2a (Method A), we inserted the FLGT sandwiched by copper blocks (cured Cu-FLGT-Cu sample, which is cured all together described in Methods) into the upper and lower heat flux meters with thermal grease applied between the surfaces of heat flux meters and the surface of copper blocks. As described in the main text, we measured $R_{int}$ and $\kappa_{TIM}$ of 5 vol.% and 10 vol.% FLGTs from the thickness-dependent $R_{TIM}$ curve ($R_{TIM}$ vs. t), using a linear fit as shown in Figure 2c. The slope of the linear fit and the intercept at y-axis are equal to $1/\kappa_{TIM}$ and $2R_{int}$, respectively.

For the $\kappa_{TIM}$ measurement on 0-2.5vol.% FLGTs as shown in Figure S2b (Method B), we mounted the disk-shape freestanding FLGT (cured alone) with thermal grease applied between two copper blocks in order to minimize the thermal interface resistance between copper blocks and the FLGT (<15mm$^2$K/W), and these samples (Cu-freestanding FLGT-Cu) were mounted between the heat flux meters again. Since the thermal interface resistance between the freestanding FLGT and copper blocks (with the grease application) is much smaller than bulk thermal resistance ($R_{TIM}=t/\kappa_{TIM}>>2R_{int}$), $\kappa_{TIM}=t/R_{TIM}$. This method is a conventional way to measure the thermal conductivity of a bulk sample when the thermal resistance ($t/\kappa_{TIM}$) of the sample is large enough, without the need of performing thickness-dependent measurements as in Method A.

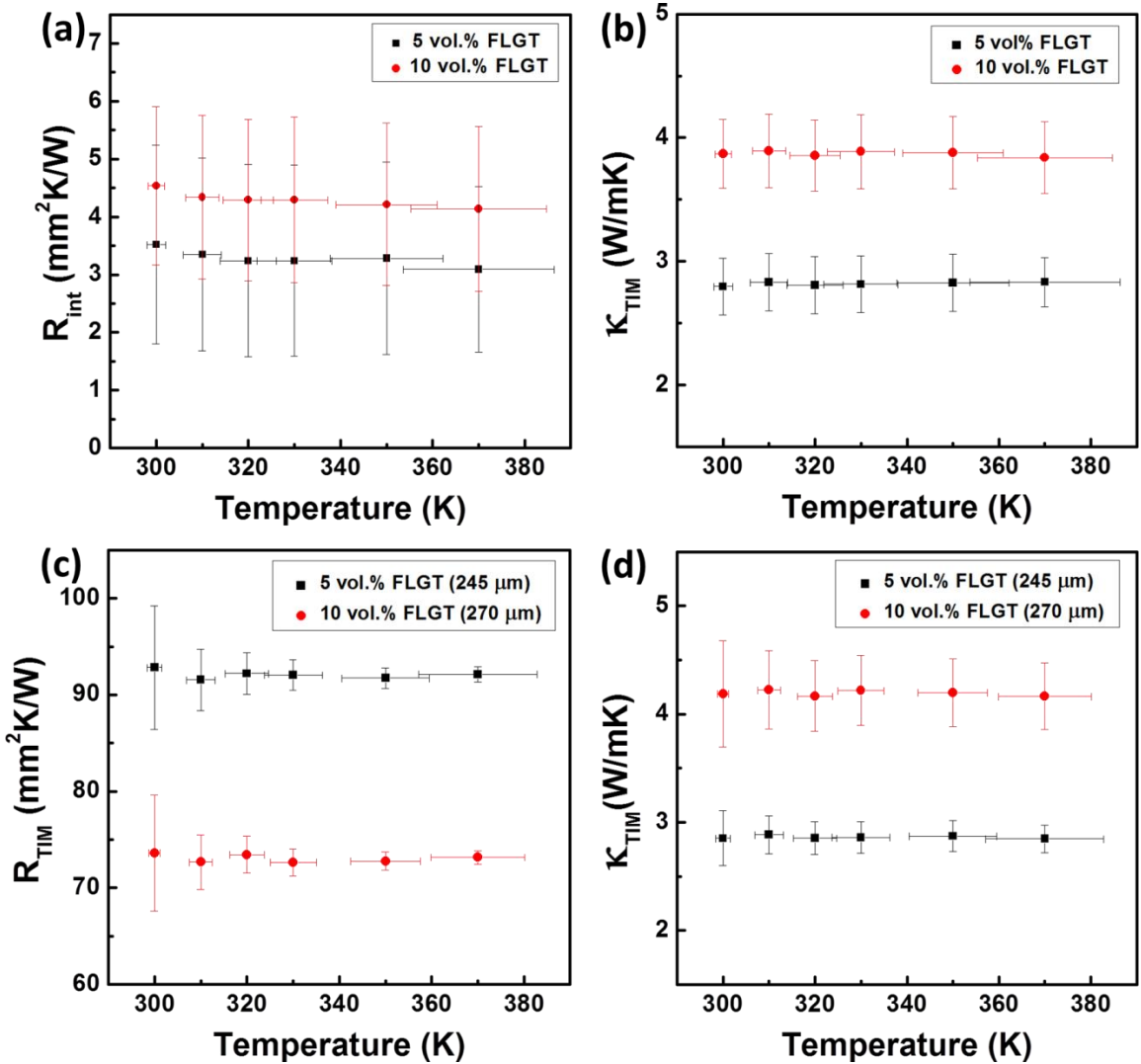

Figure S3. (a) Temperature-dependent $R_{int}$ of 5vol.% and 10 vol.% FLGTs. (b) $\kappa_{TIM}$ (inverse of the slope of the linear fit in Figure 2c) of 5 and 10 vol.% FLGTs as a function of temperature. (c) Temperature-dependent $R_{TIM}$ of 245 μm-thick 5 vol.% and 270 μm-thick 10 vol.% FLGTs. (d) $\kappa_{TIM}$ (=t/($R_{TIM}$ - 2 $R_{int}$))of 245 μm-thick 5 vol.% and 270 μm-thick 10 vol.% FLGTs as a function of temperature.

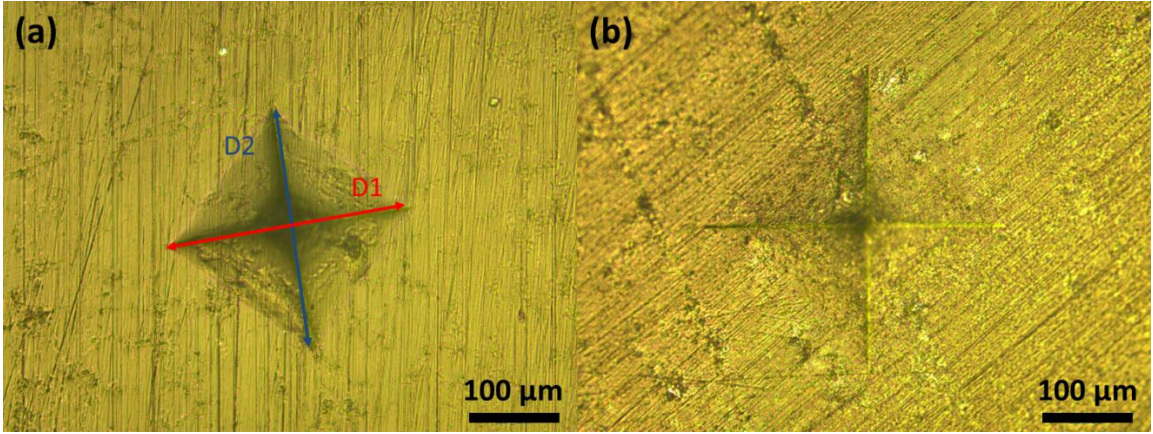

Figure S4. (a,b) Optical microscope images of representative Vickers indentations of (a) 0.25 vol.% and (b) 10 vol.% FLGTs (Hardness (in MPa) is calculated by $18.186F/((D1+D2)/2)^2$, where F is a force in kgf and D1&D2 are diagonals for indentations in mm).[1]

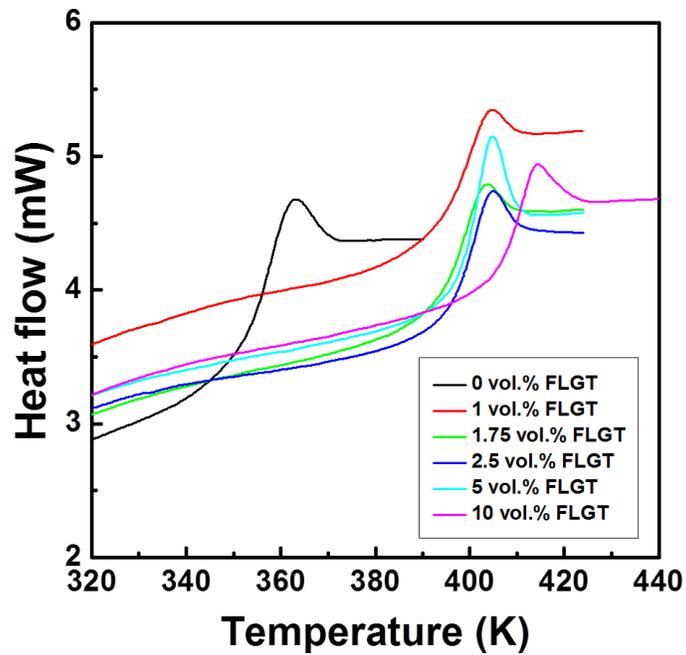

Figure S5. Differential scanning calorimetry (DSC) curves of representative FLGTs (a positive heat flow into FLGTs represents an endothermic process).

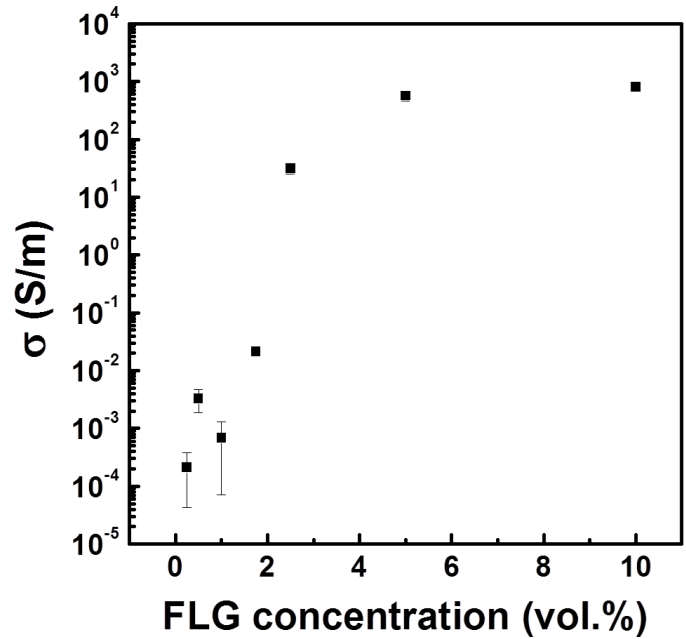

Figure S6. Electrical conductivity (σ) of FLGTs as a function of FLG concentration.

Figure S6 shows the electrical conductivity (σ) of FLGTs as a function of the FLG concentration. We observe that σ of FLGTs significantly increases by seven orders of magnitude with increasing the FLG concentration from 0.25 to 10 vol.%. It indicates that σ is dramatically enhanced after the conductive FLG networks are formed. We find that σ of our FLGTs is higher than that of commercial TIMs such as thermal greases and gap pads used as electrically insulating TIMs. In contrast to those electrically insulating TIMs, our FLGTs are more suited for applications which require high σ as well as high $\kappa_{TIM}$ and low $R_{int}$ (*e.g.* packaging for telecommunication devices required for electromagnetic shielding).

## 2. Effective medium approximation

The thermal conductivity of composites with randomly oriented ellipsoidal inclusion can be written as[2]

$$\kappa^* = \kappa_m \frac{3 + f[2\beta_{xx}(1 - L_{xx}) + \beta_{zz}(1 - L_{zz})]}{3 - f[2\beta_{xx}L_{xx} + \beta_{zz}L_{zz}]}$$

with

$$L_{xx} = L_{yy} = \frac{p^2}{2(p^2-1)} + \frac{p}{2(1-p^2)^{3/2}} \cos^{-1}p \text{ for } p < 1,$$

$$L_{zz} = 1 - 2L_{xx},$$

and

$$\beta_{ii} = \frac{\kappa_{ii}^c - \kappa_m}{\kappa_m + L_{ii}(\kappa_{ii}^c - \kappa_m)},$$

where $\kappa_m$ is the thermal conductivity of matrix, $L_{ii}$ is the geometric factor of fillers, p is the aspect ratio of fillers (p=h/L, where h is the thickness of fillers and L is the lateral size of fillers), and $\kappa_{ii}^c$ is the equivalent thermal conductivity of the filler in the composite along the ii direction. In the case of FLG fillers ($L_{xx} \approx 0$ and $L_{zz} \approx 1$ as p→0),

$$\beta_{xx} = \frac{\kappa_{xx}^c - \kappa_m}{\kappa_m} \text{ and } \beta_{zz} = \frac{\kappa_{zz}^c - \kappa_m}{\kappa_{zz}^c}.$$

The equivalent thermal conductivity of FLG encapsulated with a thin thermal barrier can be written as[3]

$$\kappa_{xx}^c = \frac{\kappa_{px}}{1 + \frac{2\alpha_K \kappa_{px}}{L\kappa_m}} \text{ and } \kappa_{zz}^c = \frac{\kappa_{pz}}{1 + \frac{2\alpha_K \kappa_{pz}}{h\kappa_m}},$$

where $\kappa_{px}$ is the in-plan thermal conductivity of FLG, $\kappa_{pz}$ is the cross-plane thermal conductivitty of FLG, and $\alpha_K (=R_B K_m)$ is the Kapitza radius. Therefore, the effective thermal conductivity ($\kappa^*$) of FLG composite TIMs (FLGTs) can be expressed as (since $\kappa_m \ll \kappa_{pz}$),

$$\kappa^* = \frac{3\kappa_m + 2f\left(\frac{\kappa_{px}}{1 + \frac{2R_B\kappa_{px}}{L}} - \kappa_m\right)}{3 - f\left(1 - \frac{\kappa_m}{\kappa_{pz}} - \frac{2R_B\kappa_m}{h}\right)} \approx \frac{3\kappa_m + 2f\left(\frac{\kappa_{px}}{1 + \frac{2R_B\kappa_{px}}{L}} - \kappa_m\right)}{3 - f\left(1 - \frac{2R_B\kappa_m}{h}\right)}.$$

**References**

(1)   Tabor, D. The Hardness of Solids. *Rev. Phys. Technol.* **1970**, *1*, 145.

(2)   Nan, C.-W.; Birringer, R.; Clarke, D. R.; Gleiter, H. Effective Thermal Conductivity of Particulate Composites with Interfacial Thermal Resistance. *J. Appl. Phys.* **1997**, *81,* 6692-6699.

(3)   Nan, C.-W.; Liu, G.; Lin, Y.; Li, M. Interface Effect on Thermal Conductivity of Carbon Nanotube Composites. *Appl. Phys. Lett.* **2004**, *85,* 3549-3551.